# AI Annotated Recommendations in an Efficient Visual Learning Environment with Emphasis on YouTube (AI-EVL)

Faeze Gholamrezaie, Melika Bahman-Abadi, and M. B. Ghaznavi-Ghoushchi

**Abstract**—In this article, we create a system called AI-EVL. This is an annotated-based learning system. We extend AI to learning experience. If a user from the main YouTube page browses YouTube videos and a user from the AI-EVL system does the same, the amount of traffic used will be much less. It is due to ignoring unwanted contents which indicates a reduction in bandwidth usage too. This system is designed to be embedded with online learning tools and platforms to enrich their curriculum. In evaluating the system using Google 2020 trend data, we were able to extract rich ontological information for each data. Of the data collected, 34.86% belong to wolfram, 30.41% to DBpedia, and 34.73% to Wikipedia. The video subtitle information is displayed interactively and functionally to the user over time as the video is played. This effective visual learning system, due to the unique features, prevents the user's distraction and makes learning more focused. The information about the subtitle text is displayed in multiple layers including AI-annotated topics, Wikipedia/DBpedia, and Wolfram enriched texts via interactive and visual widgets.

## I. INTRODUCTION

One of the most popular mass media environments is YouTube, which provides textual and audio-video information. The prevalence of instructional videos on YouTube is directly related to the number of views of that video [1]. Due to the big data, diverse topics, and the speed of rapid dissemination, discovering and tracking topics is important and challenging [2]. Websites with scientific topics have attracted many users. Among these systems, we can mention the scientific YouTube channels that receive many daily visits [5]. By extracting attributes from multimedia content (i.e., video) an information hypercube will result. In addition to video subtitles that provide textual content; each item includes a description of terms, characters, entities, places, time events, and so on.

The more relevant content in the multi-media the more understanding. Goal-oriented and efficient boosting the level of understanding of multi-media is useful and necessary for educational purposes.

On the other hand, learning is not a purely cognitive routine and the user's feeling affects learning [3]. This means the conventional experience of a user in front of the modern browser with a set of less relevant content usually disturbs the learning efficiency as well as time and bandwidth. The development of the level of education is inevitable for any society. This means the learner must be able to analyze information, present it clearly to other people in society and create innovation in expression to overcome problems [4].

To identify and track events and news, users prefer to search and watch videos on the web like YouTube. Due to the high volume of daily video releases on YouTube, one should look for an effective way to learn from it [6].

In this research, an intelligent system has been designed and developed for effective learning by searching and enriching YouTube videos. Relying on natural language processing (NLP) tools and models, the learner will experience a different preview. This is a two-folded process: one fold is the efficient bandwidth usage and the second fold is focused and NLP-based data stream.

The proposed system not only allows users not to watch less relevant content by focused and away from annoying factors such as ads, video suggestions, user comments, likes, and notifications but also gives relevant contents via AI annotations.

The AI-EVL display system works in several directions. First, before watching the video, in a smart step in a few words and visually informs the user about the contents of the video. Second, video segmentation based on time slots makes it possible to move with the help of subtitles. In the next step, at the same time as showing the video, in addition to subtitles, it also visually displays annotated information.

With the development of natural language processing tools and web application program interfaces (APIs), finding keywords in subtitles and gathering information for each entity has made significant progress in providing results and descriptions related to the video. In addition to the text, the image is used to provide information for each keyword, because the presentation of the image to convey the meaning is much more eloquent than the text [11].

There are two commonly used NLP toolkits of TextRazor and Evri for web-based text processing [17], [18]. The Evri toolkit usually comes with more computational cost than the TextRazor [19]. Therefore, this article uses TextRazor as NLP. The structure of TextRazor is such that it is possible to extract various entities such as people, time, event, and place (i.e., location) in a text. In addition to extracting entities, TextRazor can also provide connections between words. Using TextRazor increases the accuracy and speed in detecting entities [20].

The API is a useful utility in developing an interoperable multi-agent system. The YouTube Data API is used to perform search queries based on relevant keywords. Natural language processing services such as OpenCalais, Zemanta [23] and AlchemyAPI, and Google Images are used to provide information for each entity. A major problem with this structure is the use of the video title, as sometimes the video title is chosen unrelated to the content of the video [13]. For solving this problem, we used topic modeling in the AI-EVL system. The AI-EVL system provides information about video subtitle entities through APIs such as Wikipedia, DBpedia, and Wolfram. Users by selecting the video can see the topic-modeling graph about the video title and see the words that are related to the video title annotated by percent. Indexing based on a topic of interest allows for more semantic queries and access to internal information from multimedia sources.

On the other hand, to create a list of favorite educational videos with the help of web tools and natural language processing, it is possible to automate the indexing of any subject [13].

In 2020, a visual dashboard was designed and built to assist teachers in e-learning so that teachers can use the dashboard to help them quickly understand and transfer content quickly [7]. The major drawback in [7] is its limited scope of using multi-media streams, which prohibits the users from access to that content. On the other hand, unregulated access to multi-media streams is not recommended for the learning environment. At AI-EVL, we create a regulation-based learning environment that blocks any ads, over (i.e., video screen and in video popups) and video HTML page (i.e., invitations, channel info, auto-play, vulgar comments, unrelated title).

In this paper, among the feature extraction methods, the text-based method and the use of video subtitles are considered. In the first attempt to identify the entities, accurate results were provided by identifying the descriptions.

In SemWebVid structure, the descriptions of a video according to the user input information and the sound of the video lead to categorized results [8]. This structure identifies the entities of a video by creating descriptions and using the information entered by the user, and in return provides the relevant content. These services utilized provide several algorithms for identifying entities. In each of the algorithms, the ambiguities in identifying the entities are removed more accurately than the previous algorithms. Finally, among the algorithms, with the presentation of NERD [9], the existence of a sentence was well identified, and instead, the relevant content was found with the help of natural language web processing tools.

Various algorithms have been proposed to identify the entities and different parts of the sentence. Nevertheless, entity detection algorithms like NERD take a long time to deliver results. With the expansion of NLP and the emergence of pre-trained models, in addition to solving the problem of time-consuming computation, the accuracy of the results was also increased [10]. Using the experience of past work is the strength of pre-trained models. AI-EVL uses TextRazor, a pre-trained model, to be beneficial, in establish sentence structure, and entity recognition.

In another application, a schema is created by displaying a template search engine, from text, video, and music, to provide a multi-dimensional summary of the video, the structure of which leads to efficient video exploration [12]. The proposed system allows users to select smarter and more relevant results (i.e., content in the form of words) before watching a video.

To enhance the user experience when exploring video content, a framework for the display engine was introduced in [14] called RAAVE. RAAVE takes advantage of the fact that a video does not only contain single artifacts but also a combination of various temporarily restricted parallel methods (visual, audio, linguistic/textual). Current methods for displaying video content are customized for a specific case and cannot be configured to evolve user needs. To solve this problem, RAAVE works independently of the user interface by providing a display engine. RAAVE enhances the user's video exploration experience such as retrieval, navigation, and text summarization to evaluate the extent of multidimensional features and extract them from the video [14]. One of the affirmative features of a learning environment is having diverse features search and retrieve to make a better sense for the learners. In this article, by contemplation of several factors in an integrated result, it nominates an effective learning environment.

## II. RELATED WORK

One of the unique features of YouTube videos is the submission of subtitles in different languages. In the subtitle related to the film, the names of people, organizations, places, events, definitions, etc. are stated. Acquaintance with the concept of the mentioned items leads to comprehensive and complete video learning.

SemWebVid system provides related videos by receiving keywords and summaries from the user. Keywords and descriptions are automatically extracted for each video. One of the benefits of this structure is the showing of the learning paraphernalia of video in the form of graphic results [8].

Various services try to make and expand the platform for ambiguity resolving and identification of entities in a sentence. Numerous algorithms were proposed for accurate detection of Entity. In [9] it uses popular ontology services such as DBpedia ontology or YAGO [21] to classify information.

Other web tools such as DBpedia Spotlight [22], Zemanta [23], AlchemyAPI, OpenCalais, Extractive, Evri, Lupedia, Sapelo Wikimedia, and content extraction from Yahoo, was used for natural language processing [24]. During the numerous reviews among the mentioned services, some of them are more successful in identifying a certain type of entity. As you can see in Table I, some of the above-mentioned services are obsolete. The SYVSE structure uses video subtitles instead of video titles to prevent user errors in presenting results. In the steps of presenting the information and displaying it to the user, SYVSE also used the image in addition to the text, because the presentation of the image to convey the concept is much more expressive than the text [11].

TABLE I
COMPARISON OF NATURAL LANGUAGE WEB SERVICES.

| Major performance | Service status | Service name |
|---|---|---|
| Quantitative classes recognition | active | DBpedia Spotlight |
| Famous people identification | active | Zemanta |
| Individuals and cities identification | Obsolete | AlchemyAPI |
| Classification of individuals and entities and organizations | Obsolete | OpenCalais |
| Classification of countries and extraction of time | Obsolete | Extractiv |

Focusing on instructional videos, a speech video search engine called SY-E-MFSE provides keyword-based search using vocabulary, NLP tools, and the web API. It utilized a bilingual Persian English search engine [15].

Among the most widely used educational materials in academic contents such as short courses, full courses, extensions, and lectures from the leading universities, the SY-E-MFSE structure delivers user-friendly results and makes training more understandable from video content.

The RAAVE framework was introduced to enhance the experience of using video exploration. RAAVE works in three steps: extraction, indexing, and display through the matching pattern. In the first and second stages, the extracted properties are stored in the repository. It uses the existing media splitting techniques to split the video and then display the various sections in a multi-mode and customizable way. The choice of a band-specific segmentation approach depends on various factors such as the video genre [25].

A typical case on segmented algorithm is used in TED-style information videos in experiments [26]. Multi-mode feature extraction means various features that can be extracted from the video, which by displaying these contents, the video message is better conveyed to the viewer [14].

Table II provides an overall evaluation of the frameworks examined. The specified frameworks provide different results by receiving specific input from the user. In fact, in addition to differences in the type and content of input, frameworks are also different in output content. The performance of the frameworks has been inspected by considering several features such as video Search, Video display, summary, segmentation, keywords, and interactive results, relative content, video subtitle, entities labeling, and titles. The overall result of the table clearly shows the superior and user-friendly results of the AI-EVL system compared to the previous frameworks.

## III. BACKGROUND KNOWLEDGE

### A. Web API

Web-based software provides programmers with information, services, and results by providing their APIs. The web API is the interface between the web server on the server-side and the web browser on the client-side. This interface receives appropriate responses by sending requests to the server and displays them in the browser. On the one hand, using the API of different web-based software reduces the cost of programming infrastructure and, on the other hand, makes it possible to provide results in a short time [27].

### B. Named-Entity Recognition

Named-entity is equivalent to an object in the real world. Objects exist in the real world abstractly or physically. For example, a named entity can be in the category of names of people, places, events, organizations, and products. In a text containing various information, the task of identifying the Named-entity is Named-entity recognition (NER). NER places each of the entities in a predefined category. For example, it puts people, currency, and events in the category. The results of NER actions are presented to users as text enrichment or text annotation [28].

### C. Data-Driven Documents (D3)

The visual display of data contains more obvious results for the user. The web browser uses D3 to visualize data. D3 is a JavaScript library that visualizes data using SVG, HTML, and CSS. Using dynamics and visual display accelerates the transfer of concepts and makes them stay in the user's mind [29].

TABLE II
COMPARISON OF DIFFERENT ONTOLOGY FRAMEWORKS WITH AI-EVL.

| | RAAVE[14] | SemWebVid[8] | SYVSE[11] | SY-E-MFSE[15] | AI-EVL |
|---|---|---|---|---|---|
| **Video Search** | Online-TED | Online-Id | Offline | Offline-Academic | Video Search |
| **Video display** | Flexible player | Fixed frame | YouTube-like | YouTube-like | Video display |
| **Summary** | Text | Not support | Not support | Not support | Summary |
| **Segmentation** | Time-based | Not support | Not support | Not support | Segmentation |
| **Keywords** | In cloud | Subtitle | NERD | NERD | Keywords |
| **Interactive results** | Description | Links | Description | Description | Interactive results |
| **Relative content** | TED | Not support | Not support | Not support | Relative content |
| **Video subtitle** | Not support | Time-based | Time-based | Time-based | Video subtitle |
| **Entities labeling** | Not support | Not support | Tag | Tag | Video subtitle |
| **Titles** | Not support | Not support | Not support | Not support | Titles |

## IV. AI-EFFICIENT VISUAL LEARNING ENVIRONMENT APPROACH

The AI-EVL system offered in this article gets a keyword and provides a list of related videos. Enriched content is made of presented by the user by selecting a video, processing subtitles, and using the favor of artificial intelligence tools. Rich content includes titles attributed to video content.

It detects entities using a pre-trained model. The pre-training model used in this article is TextRazor. The use of this model, in addition to increasing the accuracy in identifying entities, provides better results in subject extraction, text classification, semantic discovery, and so on. Using the pre-training model, for each sentence expressed in the video, two steps are performed automatically. In the first step, the entities are extracted in the subtitle. In the second step, for each entity using the API, with the favor of web NLP tools, relevant information, descriptions, and images are extracted from various sources and the results are publicized in a suitable framework.

As shown in Figure 1, the AI-EVL system provides information about video subtitle entities through APIs such as Wikipedia, DBpedia, and Wolfram. The difference between the AI-EVL system and other research works is that this system is provided to the user in a centralized manner, which means that advertisements, video offers, channel membership offers, announcements, channel list, list playback and user activity, including comments and ratings, are blocked and not displayed to the user. Only the video and its subtitles are provided to the user in such a way that the subtitles are highlighted according to each section of the video over time and the entities of that section of the sentence are displayed to the user in the form of a network diagram. This feature is fully interactive and asynchronous. This means the learner may click on each part of the sub-title and the video will restart from that point for any required times. This will give the learner a similar sense of reading with back-and-forth moving over the pages of a book. It also supports pausing the entire recommendation system and do in-depth tasks for learning with activities like online sticky notes, re-evaluate the topic coverage, and revisit the enriched sources provided to have a clear insight of the paused content which in turn gains AI-EVL system as an effective visual learning tool. This system provides videos and related subtitles in a categorized and targeted manner away from any ads. The AI-EVL system is intelligent, meaning that it uses artificial intelligence to receive the entire subtitle and identifies important topics in the text by expressing percentages in the form of graphs.

In this paper, we propose the AI-EVL system to enhance user interaction with YouTube videos. In this system, for each video user selects, the subtitle related to the video is examined. After searching for the desired video, in the first result, the AI-EVL system visually offers subtitles for the video subject by receiving subtitles for each video. In the next, results of the AI-EVL system for each subtitle are identified and labeling is performed. In this step, we use TextRazor pre-trained model to identify the entities. The strength of the system we offer is the recognition of the existence of each subtitle sentence during the screening and execution of the video.

After identifying the entities, the next step is to enrich the content. In the enrichment step, useful and relevant information for each of the entities should be displayed to the learner. Information can be obtained from different sources for each entity.

In AI-EVL, the results of different sources per entity are presented in the most intuitive way possible. Expression of concepts in the AI-EVL system, which is done using the visual and dynamic display in the form of network diagrams; increases the user's level of understanding and retention of the content in his mind.

Figure 2 shows the layered model of the AI-EVL system. Layer 1 is the user interface that creates the graphical environment for the user to interact with the system. JavaScript is used to create visual charts, page animations, and send and receive data through Ajax. In layer 2, the subtitle sent from the user interface is imported and its entities are returned back to the user interface. This layer is also used to get information from layers 4, 5, and 6. Layer 3 is for retrieving and filtering video information that the user is searching for, and Layers 4, 5, and 6 are the same APIs used to extract information from Wikipedia, DBpedia, and Wolfram.

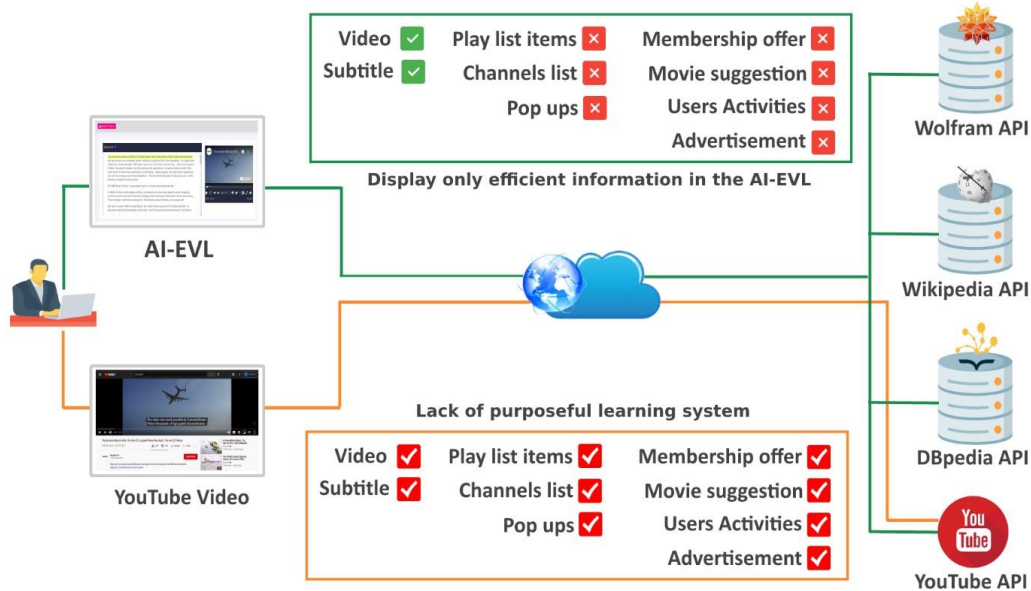

Fig. 1. Blocked and unblocked content in effective learning in AI-EVL system.

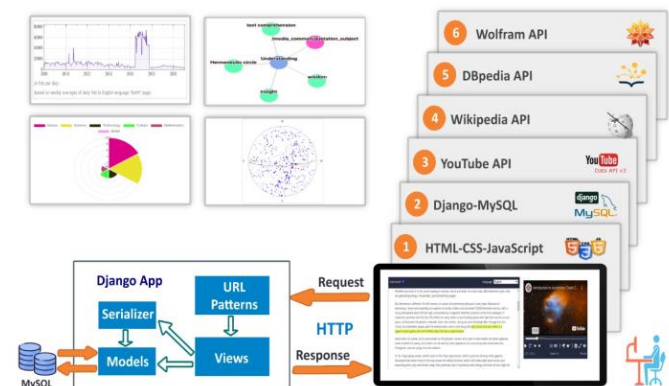

Fig. 2. AI-EVL system multi-layer architecture.

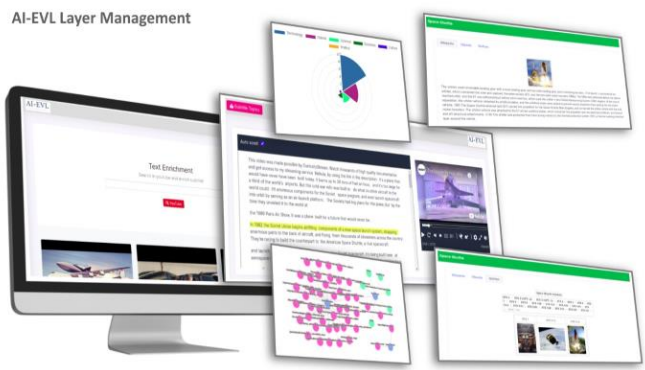

Fig. 3. Display visual layers of information in the AI-EVL system.

Figure 3 shows the general process of presenting the results of the AI-EVL system step by step. It uses Django framework by creating a layer consisting of the triple MVT (Model View Template) architecture. The top layer connects to YouTube via the YouTube Data API v3. In this connection, the keyword is sent to the YouTube server and the related videos are.

On the first page, the user have a category of related videos. For each video, the overall image, title, and duration are displayed to the user in different <div> elements. By selecting each video, the learners will have a new layer of effective visual learning. In this layer, the user learns from the educational content in an extremely focused and purposeful way. This environment is free of any advertisements and announcements of channel subscriptions, video watching, audience comments, etc.

By selecting a video, a flexible and dynamic display of the video and timed subtitles are performed. At the top of this page, a button called subtitle topics is generated. This button is for displaying smart information. The information includes smart titles related to video content. It is possible to access this smart information before watching the video. This is a unique feature. With the help of an artificial intelligence model, we extract smart titles. By viewing smart titles, it is possible to access video content. If the titles of each video are tagged, it will be possible to identify unrelated video titles. The visual display of video content in the form of a few words allows the user to select and identify videos efficiently.

During the video display, in addition to the simultaneous display of the subtitle on the left, the subtitle entities are also dynamically displayed in the form of a graphic widget. Each blue entity, in the AI-EVL system, acts as the parent of a graph. Each parent contains two types of child, related words and the remarked label. Each Child who is in the role of related words are born green. The children created under the existence label are pink. In an information delivery system, the first goal is to enhance learning.

The AI-EVL system takes a new step towards efficient and dynamic information transfer with the help of intelligent visual results. Having keywords and entities in a dynamic and flexible graphic gains the appeal and learning outcome. In the following sections, we will examine each of AI-EVL system steps and implemented algorithms in more details.

### A. Presentation of Keyword-Related Videos

On the AI-EVL system homepage, the keyword is first entered by the user. Next, with the help of YouTube API, we will receive YouTube information. Via the YouTube API V3, by sending the keyword, YouTube sends the videos information related to the keyword as output. YouTube API assigns a KEY API to each user to access YouTube information. After receiving the API KEY, the keyword entered by the user will be sent as a request. YouTube returns the ID number of the related videos in response to the request. For each video ID, a new request is sent to YouTube to provide the URL of the video associated with the ID. In return for each video ID, YouTube returns many items, including the video title, duration, video description, and more.

### B. Getting Video Subtitles

Since in the AI-EVL system, the subtitle of each item is considered as input, we must receive the subtitle of each video. The AI-EVL system, by displaying video subtitles in a separate HTML element, offers additional features including display the title of each paragraph, marking the subtitle text to match the video time, and synchronising text and video in an interactive manner.

### C. AI-based Title Suggestion and Avoiding Inefficient Surfing

Since one of the surfing challenges on YouTube is the less consistency of the video title related to its content [13]. To overcome this problem, in the AI-EVAL prior to watching the video by the user there is a facility allows to decide whether watch or not, an individual video based on the AI annotation on the video title. This gains a pre-dominant smart model, instead of conventional human-oriented look and evaluate.

Providing intelligent video content results not only saves time in selecting the desired video but also plays a role in categorizing videos. The output of the smart model is a set of smart titles. Transferring video content in the form of titles leads the user to make the right choice in learning efficiently.

### D. Extraction of Subtitle Related Entities

AI-EVL system makes it possible to extract the entities of each sentence from the subtitle. To properly extract the entities of a sentence, the various components of the sentence must be well-identified. After identifying the components of the sentence, with the help of NLP tools, the entities of each sentence are extracted using the pre-trained TextRazor model. The use of pre-trained models makes it possible to use previous experiences in a variety of contexts to improve the learning experience.

### E. AI-EVL Basic and NLP Procedures

The pseudocode of implemented AI-EVL and NLP procedures are presented in Algorithms 1 and 2 respectively. The former receives videos related to inquiries. By selecting the desired video, the user enters an effective visual learning environment. For each highlighted subtitle, entity-related text and ontology entities are extracted. The ontology of any entity includes labels and synonyms. The results associated with each entity are calculated in latter and then sent back to the former for display information in the learning environment.

---

**Algorithm 1** AI-EVL procedure

**Input:** keyword, unique-key

1: response = Post a keyword request to YouTube by unique-key
2: **for** $iteration = 1, 2, \ldots, len(Response)$ **do**
3:    Show video-preview and video-title and time
4: **end for**
5: **if** select video **then**
6:    Move to learning environment AI-EVL
7:    Show visual smart titles
8:    Show video and highlight subtitles and ontology environment
9: **end if**
10: **if** subtitle highlighted **then**
11:    Show Web NLP Procedure results in ontology
12: **end if**

---

**Algorithm 2** Web NLP procedure

**Input:** entity, unique-key
**Output:** label, synonym, description

1: Get response from pre-trained model (entity)
2: label ← response.label
3: synonym ← response.Related
4: description ← request (entity) + unique-key

---

## V. EVALUATION OF THE AI-EVL FRAMEWORK

By considering the search results, more intelligent results can be extracted. Smart results come from search terms. Every year, Google provides Google trends based on the words that users have searched for.

In this article, Google Trend 2020 data was considered as an AI-EVL input database. By receiving each of these words, AI-EVL first presents the related videos and then, by selecting the most relevant video, ontology the related word. Because of the popularity of these words in 2020 and the release of videos in 2020, it can be said that AI-EVL examines each phrase in Google Trend within a video. The rich results presented for each phrase examine people's views and mindsets more closely and provide tightly results. The Google Trends data [16] is categorized in 11 sections: Searches, News, Actors, Athletes, Games, Loss, Lyrics, Videos, People, Recipes, and TV Shows. Each section contains the top 10 phrases. For example, in

Searches, the top 10 words are *Coronavirus, Election result, Kobe Bryant, Zoom, IPL, India vs New Zealand, Coronavirus update,* and *Coronavirus symptoms, Joe Biden, Google Classroom,* respectively.

According to Table III and the above words, the percentage of ontology information source has been calculated for each group.

According to the color aggregation of each group in Figure 4, it can be seen which source is more active in providing information in related area. For data from the *Searches* and *Videos* groups, most of the data is at the level of DBpedia and Wikipedia. On the other hand, more *Searches-related* retrieved data from the WolframAlpha section is observed. Most of the data from *News, Actors, Athletes, Loss, Lyrics, People,* and *Recipes* are distributed in all three dimensions WolframAlpha, DBpedia Wikipedia. Games group are mostly distributed in two dimensions: Wikipedia and DBpedia. TV Shows group data is present in three dimensions at the same time but is mostly in Wikipedia and DBpedia. Based on these results, if the google trends words 2020 apply to the AI-EVL system; each keyword has a good source of enrichment from three sources: Wikipedia, DBpedia, and Wolfram, which provide valuable information to the user.

Table IV shows the distinct and preferred performance of the AI-EVL framework over previous ontology frameworks.

The AI-EVL system acts as a filter to prevent the display of miscellaneous information such as hateful, criminal, racist, narcotics, and inappropriate content for example. This protection takes place intelligently, which is one of the purposes of teaching and learning supervised recommendation. Assuming a special module next to it in the discussion of advice-based education, this system acts in a way that prevents the publication of non-video-related content and acts as a parental-control for students.

TABLE III
PERCENTAGE OF RESULTS PROVIDED BY KEYWORDS PER GROUP.

| Groups | Wikipedia | DBpedia | Wolfram |
|---|---|---|---|
| Searches | 35.7 | 14.3 | 50 |
| News | 36.85 | 26.3 | 36.85 |
| Actors | 37.05 | 33.35 | 29.6 |
| Athletes | 32.15 | 32.15 | 35.7 |
| Games | 52.94 | 41.17 | 5.89 |
| Loss | 33.33 | 33.33 | 33.33 |
| Lyrics | 33.33 | 27.78 | 38.89 |
| Movies | 47.62 | 9.52 | 42.86 |
| People | 34.48 | 31.04 | 34.48 |
| Recipes | 34.48 | 34.48 | 31.04 |
| TV Shows | 45.46 | 31.82 | 22.72 |

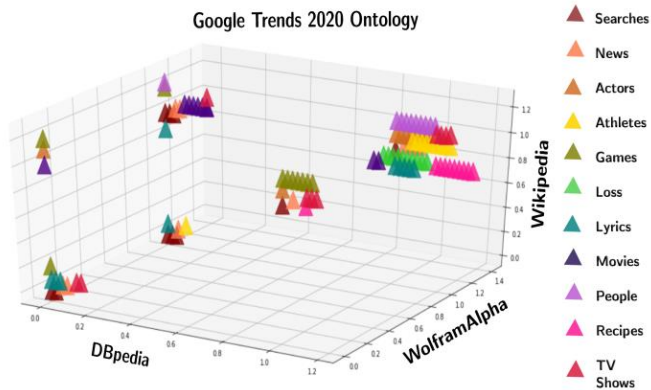

Fig. 4. 3D display of Google Trend data ontology results in 2020

TABLE IV
COMPARISON OF DIFFERENT ONTOLOGY FRAMEWORKS WITH AI-EVL.

| | RAAVE[14] | SemWebVid[8] | SYVSE[11] | SY-E-MFSE[15] | AI-EVL |
|---|---|---|---|---|---|
| **Online- Matches with YouTube search** | Search offline between academic lectures. | Search offline between videos. | Search online between videos. | Online- about TED movies | Video Search |
| **Subtitle-based-Support different language, No extra contents** | View all the features of a video player. | View all the features of a video player. | View movies in a fixed frame. | Display video with resizable capability. | Video display |
| **Graphical chart-User friend** | No content | No content | No content | Provide a summary of the video | Summary |
| **Time-based, Highlight-based** | No content | No content | No content | Segment the videos and present them with the relevant time | Segmentation |
| **Text-razor** | Provide a description for each entity of the NLP tool | Extraction of entities with the help of NERD | Extract keywords in subtitles | Keyword extraction and cloud display of keywords | Keywords |
| **Network graph-Visual objects** | Provide description of entity. | Provide a description of entity. | Provide keyword descriptive links. | Provide descriptions based on keywords. | Interactive results |
| **Wikipedia** | No content | No content | No content | Suggest related videos in TED's lectures. | Relative content |
| **Flexible, Clean-Interactive** | Show video subtitles for each time slot | Show video subtitles of time slot. | Show video subtitles of time slot. | No content | Video subtitle |
| **Flexible, Clean-Interactive** | Show video subtitles of time slot. | Show video subtitles of time slot. | Show video subtitles of time slot. | No content | Video subtitle |
| **AI-based** | No content | No content | No content | No content | Titles |

## VI. CONCLUSION

AI-EVL provide a well-formed scientific platform away from any ads and inappropriate content based on YouTube videos. In fact, it provides smart and graphical results, allowing user to learn from YouTube videos in a secure environment. With the help of artificial intelligence, effective results are obtained that are not obvious at first glance. In this environment, the user has rich graphic results along with the video. Rich results help the user to learn topics, concepts, events, places, etc. The AI-EVL system enables learning from YouTube videos in a safe environment by providing intelligent and graphical results. The presented platform covered several cases so that the learning platform can be used for all ages and all topics. This system, utilizes an AI-Assisted pre-recommender tool chain in a multi-layered web-based application. Artificial intelligence, helps the user to select the more appropriate video and provides him/her with an effective visual learning platform away from any less-relevant contents.

## REFERENCES

[1] I. Duncan, L. Yarwood-Ross, and C. Haigh, "YouTube as a source of clinical skills education," *Nurse Education Today*, vol. 33, no. 12, pp. 1576–1580, Dec. 2013, doi: 10.1016/j.nedt.2012.12.013.

[2] Z. Lu, Y.-R. Lin, X. Huang, N. Xiong, and Z. Fang, "Visual topic discovering, tracking and summarization from social media streams," *Multimedia Tools and Applications*, vol. 76, no. 8, pp. 10855–10879, Sep. 2016, doi: 10.1007/s11042-016-3877-1.

[3] Z. Karamimehr, M. M. Sepehri, and S. Sibdari, "Automatic Method to Identify E-Learner Emotions Using Behavioral Cues," *IEEE Transactions on Learning Technologies*, vol. 13, no. 4, pp. 762–776, Oct. 2020, doi: 10.1109/tlt.2020.3020497.

[4] N. Songkram, "Virtual smart classroom to enhance 21st century skills in learning and innovation for higher education learners," *2017 Tenth International Conference on Mobile Computing and Ubiquitous Network (ICMU)*, Oct. 2017, doi: 10.23919/icmu.2017.8330109.

[5] S. Rosenthal, "Motivations to seek science videos on YouTube: free-choice learning in a connected society," *International Journal of Science Education, Part B*, vol. 8, no. 1, pp. 22–39, Sep. 2017, doi: 10.1080/21548455.2017.1371357.

[6] J. Sang and C. Xu, "Browse by chunks," *ACM Transactions on Multimedia Computing, Communications, and Applications*, vol. 7S, no. 1, pp. 1–18, Oct. 2011, doi: 10.1145/2037676.2037687.

[7] I. Amarasinghe, D. Hernandez-Leo, K. Michos, and M. Vujovic, "An Actionable Orchestration Dashboard to Enhance Collaboration in the Classroom," *IEEE Transactions on Learning Technologies*, vol. 13, no. 4, pp. 662–675, Oct. 2020, doi: 10.1109/tlt.2020.3028597.

[8] T. Steiner and M. Hausenblas, "SemWebVid - Making Video a First Class Semantic Web Citizen and a First Class Web Bourgeois," in *Open Track of the Semantic Web Challenge*, 2010, vol. 658, no. 4, pp. 97–100.

[9] G. Rizzo and R. e. Troncy, "NERD: A Framework for Unifying Named Entity Recognition and Disambiguation Extraction Tools," in *Proceedings of the 13th Conference of the European Chapter of the Association for Computational Linguistics*, 2013, pp. 73–76.

[10] A. Radford, K. Narasimhan, T. Salimans, and I. Sutskever, "Improving Language Understanding by Generative Pre-Training," in *OpenAI*, 2018, p. 12.

[11] B. Farhadi and M. B. Ghaznavi-Ghoushchi, "Creating a novel semantic video search engine through enrichment textual and temporal features of subtitled YouTube media fragments," in *ICCKE 2013*, Oct. 2013, pp. 64–72, doi: 10.1109/iccke.2013.6682857.

[12] F. A. Salim, F. Haider, O. Conlan, and S. Luz, "An approach for exploring a video via multimodal feature extraction and user interactions," *Journal on Multimodal User Interfaces*, vol. 12, no. 4, pp. 285–296, Jul. 2018, doi: 10.1007/s12193-018-0268-0.

[13] B. IANCU, "Web Crawler for Indexing Video e-Learning Resources: A YouTube Case Study," *Informatica Economica*, vol. 23, no. 2/2019, pp. 15–24, Jun. 2019, doi: 10.12948/issn14531305/23.2.2019.02.

[14] F. Salim, "An Alternative Representation of Video via Feature Extraction (RAAVE)," Ph.D. dissertation, Dept. CS., Trinity Univ., Dublin, Ireland, 2019.

[15] B. Farhadi, "Creating a Semantic Academic Lecture Video Search Engine via Enrichment Textual and Temporal Features of Subtitled YouTube EDU Media Fragments," *International Journal of Computer Applications*, vol. 96, no. 13, pp. 13–18, Jun. 2014, doi: 10.5120/16853-6719.

[16] "Google's Year in Search," 2021. Accessed on: Mar. 21, 2021. [Online]. Available: https://trends.google.com/trends/yis/2020/MY/

[17] "Evri.com domain is for sale | Buy with Epik.com," Accessed on: Apr. 03, 2021. [Online]. Available: https://www.evri.com/

[18] L. Chen, M. Benedikt, and E. Kharlamov, "QUASAR: Querying Annotation, Structure, and Reasoning," *Proceedings of the 15th International Conference on Extending Database Technology - EDBT '12*, pp. 618–621, 2012, doi: 10.1145/2247596.2247680.

[19] "TextRazor - The Natural Language Processing API," 2009. Accessed on: Dec. 15, 2019. [Online]. Available: https://www.textrazor.com

[20] Y. Ju, B. Adams, K. Janowicz, Y. Hu, B. Yan, and G. McKenzie, "Things and Strings: Improving Place Name Disambiguation from Short Texts by Combining Entity Co-Occurrence with Topic Modeling," *Lecture Notes in Computer Science*, pp. 353–367, 2016, doi: 10.1007/978-3-319-49004-5_23.

[21] "Yago," 2012. Accessed on: Apr. 03, 2021. [Online]. Available: https://www.yago-knowledge.org

[22] "DBpedia Spotlight - Shedding light on the web of documents," Accessed on: Apr. 03, 2021. [Online]. Available: https://www.dbpedia-spotlight.org

[23] "Zemanta I Programmatic Ad Technology Built for Engagement," Accessed on: Apr. 03, 2021. [Online]. Available: https://www.zemanta.com

[24] "APIs - Yahoo Developer Network," Accessed on: Mar. 21, 2021. [Online]. Available: https://developer.yahoo.com/api/

[25] M. Bouyahi and Y. B. Ayed, "Video Scenes Segmentation Based on Multimodal Genre Prediction," *Procedia Computer Science*, vol. 176, pp. 10–21, 2020, doi: 10.1016/j.procs.2020.08.002.

[26] Haojin Yang and C. Meinel, "Content Based Lecture Video Retrieval Using Speech and Video Text Information," *IEEE Transactions on Learning Technologies*, vol. 7, no. 2, pp. 142–154, Apr. 2014, doi: 10.1109/tlt.2014.2307305.

[27] M. Grahl *et al.*, "Archive WEB API: A web service for the experiment data archive of Wendelstein 7-X," *Fusion Engineering and Design*, vol. 123, pp. 1015–1019, Nov. 2017, doi: 10.1016/j.fusengdes.2017.02.047.

[28] M. Marrero, J. Urbano, S. Sánchez-Cuadrado, J. Morato, and J. M. Gómez-Berbís, "Named Entity Recognition: Fallacies, challenges and opportunities," *Computer Standards & Interfaces*, vol. 35, no. 5, pp. 482–489, Sep. 2013, doi: 10.1016/j.csi.2012.09.004.

[29] M. Bostock, V. Ogievetsky, and J. Heer, "D³ Data-Driven Documents," *IEEE Transactions on Visualization and Computer Graphics*, vol. 17, no. 12, pp. 2301–2309, Dec. 2011, doi: 10.1109/tvcg.2011.185.

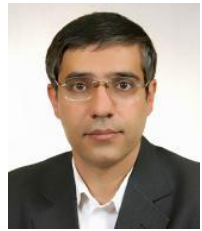

**M. B. Ghaznavi-Ghoushchi** received a B.Sc. degree from the Shiraz University, Shiraz, Iran, in 1993, the M.Sc., and Ph.D. degrees both from the Tarbiat Modares University, Tehran, Iran, in 1997, and 2003 respectively. During 2003-2004, he was a researcher at TMU Institute of Information Technology. He is currently an Associate Professor with Shaded University, Tehran, Iran. His interests include VLSI Design, Low-Power and Energy-Efficient circuit and systems, Computer-Aided Design Automation for Mixed-Signal, and UML-based designs for SOC and Mixed-Signal.

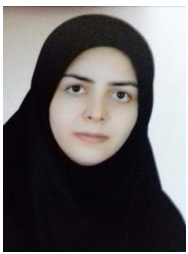

**Faeze Gholamrezaie** received a bachelor's degree in computer science from Shahed University, Tehran, Iran, in 2019. She is currently a graduate student in Soft Computing and Artificial Intelligence. She has been a teaching assistant at Shahed University since 2019. Her research interests focus on machine learning and deep neural networks to improve textual interactions.

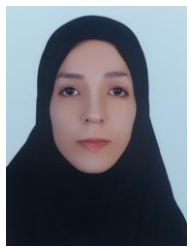

**Melika Bahman-Abadi** received a B.Sc. degree in Computer-Software Engineering from Dr. Shariaty Technical and Vocational College, Tehran, Iran, in 2019 and is an M.Sc. student of Computer Science-Data Mining at Shahed University, Tehran, Iran from 2019. Since 2018, she has worked in the field of full-stack web development. She is currently working on cyberspace data analysis, data visualization, and signal processing.